# Towards an Adaptive and Normative Multi-Agent System Metamodel and Language: Existing Approaches and Research Opportunities

Marx Viana, Carlos Lucena
*Laboratory of Software Engineering (LES)*
*Pontifical Catholic University of Rio de Janeiro (PUC-Rio)*
Rio de Janeiro, Brazil
(mleles, lucena) @inf.puc-rio.br

Paulo Alencar
*David R. Cheriton School of Computer Science*
*University of Waterloo (UW)*
Waterloo, Canada
palencar@cs.uwaterloo.ca

***Abstract*—Several Multi-Agent System (MAS) metamodels and languages have been proposed in the literature to support the development of agent-based applications. MAS metamodels are used to capture a collection of concepts the relevant entities and relationships in the MAS domain, which include entities such as agent, message, role, action and plan, and relationships that represent, for example, that a role is responsible for one or more tasks. In addition, to models, MAS modeling languages have also been used to support the development of MASs in a wide variety of domains, including social networking, robotics, security and smart city environments. However, there is a lack of support in these models and languages for abstractions involving norms and adaptations as well as their interactions. This paper presents a survey of some existing metamodels and languages and compares their expressiveness using abstractions related to agents, norms and adaptation. The comparison serves as a basis for the definition of a new MAS metamodeling.**

*Keywords: Multi-agent sytems, metamodels, modeling languages, adaptation, norms.*

## I. Introduction

Multiagent Systems (MASs) are systems composed of autonomous entities called agents, and in which each agent has its own goals and constraints [1] [2] [3]. Because these highly dynamic systems often involve complex and non-deterministic behavior based on *learning*, *norms self-adaptation* and *self-organization*, they are difficult to understand and represent. As a result, a major research challenge in MAS research is how to model and develop these systems, especially those that have higher degrees of autonomy.

To address this challenge, several of MAS metamodels and languages have been proposed in the literature. MAS metamodels are used to capture a collection of concepts the relevant entities and relationships in the MAS domain, which include entities such as agent, message, role, action and plan, and relationships that represent, for example, that a role is responsible for one or more tasks. In addition, MAS modeling languages have also been used to facilitate the development of MASs in a wide variety of domains, including social networking, robotics, security and smart city environments.

However, there is a lack of support in these models and languages for abstractions involving norms and adaptations as well as their interactions. This paper presents a survey of existing metamodels and languages and compares their expressiveness using abstractions related to agents, norms and adaptation. The comparison serves as a basis for the definition of a new MAS metamodeling approach called ANA-ML.

The remainder of this paper is organized as follows. Section II presents some background related to software agents, and adaptive agents, and normative systems. Section III describes some existing MAS metamodels and modeling languages, and research opportunities that arise as existing approaches are compared with respect to abstractions such as norms and adaptation. Finally, Section IV our conclusion and future research prospects.

## II. Background

This section introduces the main concepts needed to understand this work, including the concepts related to software agents, adaptive behaviour, normative systems, and MAS metamodels.

### A. Software Agents and the BDI Architecture

Software agents are autonomous entities able to perform tasks without human intervention [3]. A software agent is capable of transforming its own environment, and therefore, it needs to take into consideration the state of the environment, as well as its mental state, in order to understand its behavior. The mental state of an agent is composed of its state and behavior at a given time, i.e., goals, beliefs, decisions and intentions linked to its plans and actions. When it executes actions, the agent can change its mental state, introducing new perceptions about the environment, and by sending and receiving messages from other agents.

The agent behavior is characterized based on its plans, actions and environment norms, which influences the agent autonomy and interactions [2]. Nevertheless, software agents can send and receive messages in the environment of other agents. The agent autonomy is given through its capacity of acting proactively. Finally, agents are adaptive entities, that may adapt its state and behavior regarding changes and restrictions in the environment.

Due to the complexity associated with the development of multi-agent systems, which typically involves thread control, message exchange across the network, cognitive ability, and discovery of agents and their services, several architectures and platforms have been proposed. One of the widely known architectures for designing and implementing cognitive agents is the belief-desire-intention (BDI) architecture [2], following a model initially proposed by Bratman [4], which consists of beliefs, desires and intentions as mental attitudes that influence human action. The basic assumption of the BDI model is that actions are derived from a process called practical reasoning, which consists of two steps. In the first step, called deliberation (of objectives), one makes the selection of a set of desires that must be reached, according to the current state of the agent's beliefs. The second step is determining the desires produced as a result of the previous step [5]. The three mental states that make up the BDI model are detailed below:

- Beliefs - represent the characteristics of the environment, which are updated appropriately after each action is perceived. These beliefs represent knowledge about the world;
- Desires - contain information on the objectives to be achieved, and the priorities and costs associated with the various objectives. They can be thought of as representing the motivational state of the system;
- Intentions - represent the current action plan chosen. Capture the deliberative component of the system.

Rao and Georgeff [2] have adopted the BDI model for software agents and have presented a formal theory and an abstract BDI interpreter that is the basis for current BDI systems. The interpreter operates on the agent's beliefs, goals and plans. The main task of the interpreter is to implement a process that selects and executes plans for a given objective or event.

The BDI agent architecture is presented in Fig. 1. There are seven major components in this arechitecture:

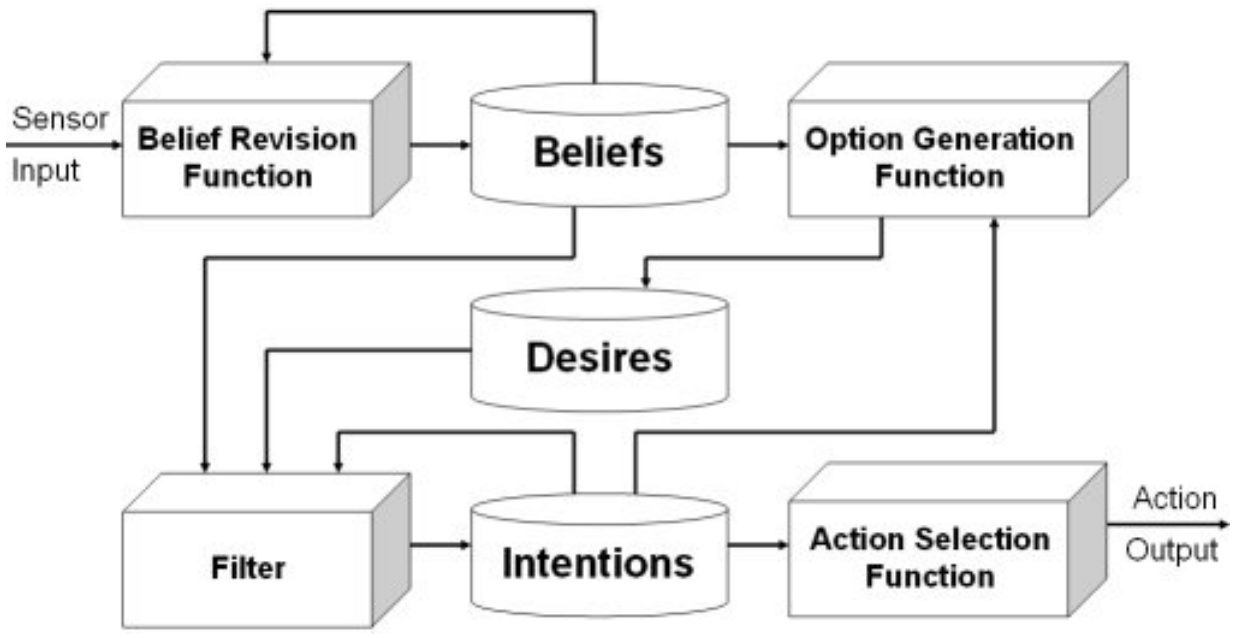

Fig. 1. Generic BDI Architecture [3].

- a set of current beliefs, representing the agent's information about its environment;
- a belief revision function, which determines a new set of beliefs from the perceived sensor input and beliefs of the agent;
- an option generation function, which determines the options available to the agents (their desires), based on their beliefs about their environment and their intentions;
- a set of current desires that represent the possible action plans available to the agent;
- a filter function, which represents the agent's deliberation process, which determines the agent's intentions based on its current beliefs, desires and intentions;
- a current set of intentions, which represents the agent's current focus, that is, those states that the agent is determined to achieve;
- an action selection function, which determines the action to be performed based on the agent's current intentions.

### B. Adaptive Agents

The term adaptation means ”any change in the structure or functioning of an organism that makes it better suited to its environment” [6]. Software adaptation refers either to the software being adapted or the evolutionary process leading to the new adapted software. More specifically, in the context of MASs, adaptations can be observed at two levels, which involve structural and behavioral changes. These changes aim at making applications more adaptable to their evolutionary context [7].

Evolutionary or adaptive changes that happen at design or compile time and are intended to deal with predictable situations such as software change requests. On the other hand, software can also change and adapt to deal with unpredictable changes in the operating software environment. A promising solution for software adaptation is introducing self-adaptive software systems that can manage change dynamically while running efficiently and reliably [8]. According to [9], self-adaptation systems are able to modify their behavior or structure in response to their perception of the environment, the system itself, and their goals. One of the main advantages of self-adaptive software is its ability to manage the complexity arising from highly dynamic and non-deterministic operating environments.

For a software system to be considered self-adaptive [10], it should support a specific set of features, which include having: (i) the ability to observe changes in its operating environment; (ii) the ability to detect and diagnose changes in the operating environment and assess its own behavior; (iii) the ability to change its own behavior to adapt to new changes; and (iv) support for dynamic behavior (i.e. its internal and external behavior should be able to change intentionally and automatically). Even though many approaches [8], [9], [10] describe systems' self-adaptation, they do not focus on modeling and implementing software agents. Therefore, although properties related to self-adaptive systems are considered important (i.e. autonomy, reasoning, proactivity), they are not explicitly addressed by existing approaches [17], especially when both adaptation and norms have to be considered.

### C. Normative Systems

Norms are mechanisms that enable an agent to require other agents behave in a certain way [11]. In other words, norms can regulate the agents' behavior by representing the way in which an agent understands the responsibilities of others [12]. In a normative system, the software agents work under the belief that other agents will behave according to prescribed norms. The use of norms is a precondition in MASs in which the members are autonomous, but not self-sufficient and, therefore, cooperation is required and needs to be assured by specific mechanisms to support it.

TABLE I
PROPERTIES OF NORMS

| Property | Description |
|---|---|
| Addressee | It is used to specify the agents or roles responsible for fulfilling a norm. |
| Activation | It defines the activation condition for a norm to become active. |
| Expiration | It defines the expiration condition for a norm to become inactive. |
| Rewards | It is used to represent the set of rewards given to an agent for fulfilling a specific norm. |
| Punishments | It is the set of punishments given to an agent for violating a norm. |
| Deontic Concept | It is used to indicate if a norm states an obligation, a permission or a prohibition. |
| State | It is is used to describe the set of states being regulated. |

Moreover, norms can be used to achieve different purposes, ranging from the construction of a simple agreement between software agents to more complex cases involving legacy systems [13]. Therefore, different aspects can be used to characterize norms. Table I presents the properties of norms. First, norms are always created to be obeyed by a set of agents to achieve specific social goals. Second, norms are not always applied, and their activation conditions depend on the context of the software agents involved in a specific interaction. The type of entity a norm regulates must be known, and it can be an agent's action or the state of the environment. Finally, in some cases, norms may provide a set of sanctions to be imposed when agents fulfill or violate them [13], [12], [11].

Nevertheless, norms and adaptation are indispensable to MAS models in both internal and external design and run times. Many types of practical MAS applications (i.e. [14], [15], [16]) require support for adaptive and normative mechanisms at design time and runtime.

## III. Existing MAS Modeling Approaches

Having described relevant background, in this section we discuss works related to (i) metamodels and (ii) modeling languages, both in the context of MAS systems. The literature reports very distinct and varying sets of abstractions suitable for different domains [17], [18], [19], [16], [14], [20], [21]. Each methodology has incorporated its own abstractions for modeling the different multi-agent systems concepts, and there is no agreement about a common group of abstractions that can be used across different methodologies. Several authors recognized the importance of modeling agents in their environment both in design time and in runtime [14], [22], [23]. In general, these authors also observed that most modeling languages do not represent some important concepts that are present in MASs, such as adaptive behavior and norms [14], [15]. Nevertheless, norms and adaptation are relevants and indispensable to the software agent's internal and external design and run times.

### A. MAS Metamodels and FAML

Several MAS metamodeling approaches have been introduced in the literature [14], [24], [22]. A metamodel involves a collection of concepts such as things and terms associated with a domain, which basically include relevant entities and relationships. In the case of MAS metamodels entities are, for example, agent, message, role, action, and plan. Instead of describing specific metamodels individually, we describe in this section a generic metamodel called FAML.

FAML [14] is a process-independent metamodel for an agent-oriented modeling language. It allows the description of software components of any multiagent system, as it captures problem-independent concepts (e.g., agents, resources) involved in multiagent system requirement description and system design with different abstractions. FAML's suitability for supporting modeling language development is demonstrated through a comparison with existing methodology-specific metamodels.

The FAML metamodel is constructed by a combination of bottom-up and top-down analysis and best practices. The expressiveness of the FAML metamodel concepts and their relationships were evaluated in [12] in comparison with two agent-oriented metamodels, namely TAO [25] and Islander [26].

The FAML metamodel has two layers: design-time and runtime layers. Each layer may have two scopes: system-related or agent-related. The design-time layer captures entities and relationships related to the "system as developed" and the runtime layer captures those related to the "system as being executed".

To give an idea about how the metamodel looks like, we present in Fig. 2 some of the design-time classes. At design-time, the concept of System plays a central role that satisfy the functional or nonfunctional requirements. Task are derived from requirements. Design-time entities include Agent, AgentDefinitions, Task, MessageSchema, and System and design-time relationships include CanUse between Role and MessageSchema, and ResponsibleFor between Role and Task.

However, concepts such as adaptation and norms are not included in the metamodel structure, neither at design nor execution times. In doing so, the agents cannot reason about the norms nor adapt to changes in the environment.

Cernuzzi [27] provided a comprehensive description of the GAIA, a methodology initially developed to model small-scale multiagent systems [3]. After undergoing several modifications [28], the authors seek to show the advantages of using GAIA to create models in the FAML social and agent level [14]. In addition, various efforts have been made to model organizational structures and the norms governing the agents' overall behavior in the organization. Even after the modifications to the GAIA methodology, it still lacks support to model how the agent understands those norms and how it deals with them. The authors mainly addressed the relation between the norms and the roles of the agents in the organization. Further, this methodology does not consider possible adaptations of the agent to deal with norms in the environment; we introduce these adaptations in our modeling language.

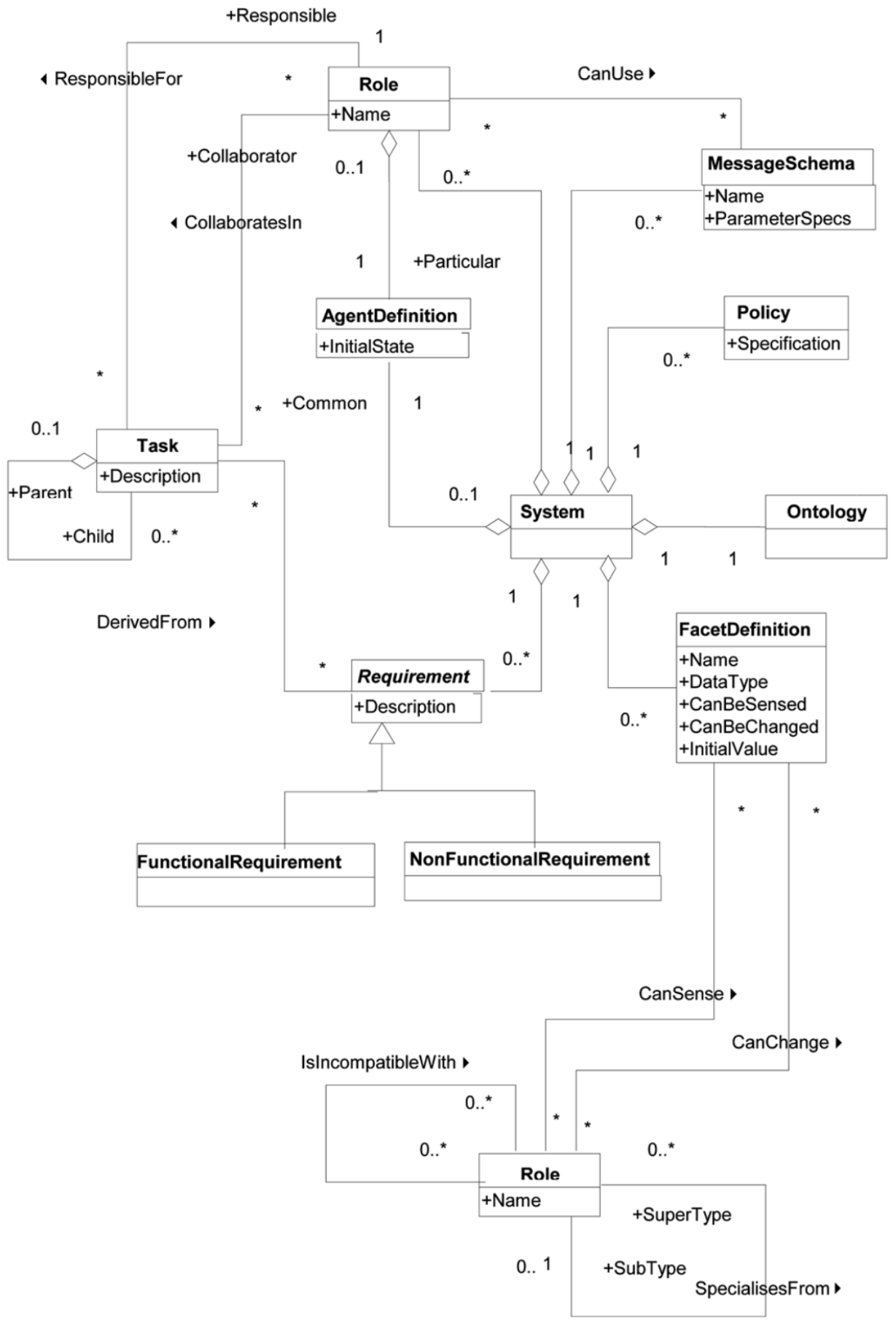

Fig. 2. Partial FAML Metamodel ( [14]).

In contrast, the metamodels presented in this section do not include a description of the internal architecture of normative agents and how they adapt their behaviors to deal with norms.

### *B. MAS Modeling Languages*

Several languages have been proposed for the modeling of MASs ( [3], [28], [24], [29], [20], [14], [30], [31], [32], [33]. Each of these approaches will be described in the following paragraphs.

Bernon [33] proposed the ADELFE methodology, which is capable of guiding and helping designers to build adaptive MASs. An important point in this theory is that the agents are not adaptive; instead, the whole system is. ADELFE provides an emerging global function of the system, while a local function is provided by agents. The global function is classified as emergent because it is not encoded within the agent. Agents are not aware of this global function. The overall function of this system results from the collective behavior of the different agents that compose it. Although the adaptation of the system is carried out globally, the ADELFE methodology does not deal with the norms in the system. Furthermore, since the agent does not adapt individually, the approach is not adequate to model normative adaptive agents.

ADELFE 3.0 [34] can be used to design non-collaborative situations an agent can find or create. For each of these situations, an agent must leave the actions to be performed and ensure that the agent resumes these actions while staying in a state of cooperation with other agents and with itself. The authors of this paper adopted a simulation-based design approach to assist design in detecting and correcting non-cooperative situations. Although [34] created a self-designing function in agents, it was never shown how the atomic characteristics of an agent were connected to the concepts of adaptation. In addition, the authors did not consider the agents' dealings with the norms while keeping their autonomy. Finally, the authors modeled non-collaborative situations in which the agents are restricted to behave in a cooperative way, thus removing part of the agents' autonomy.

Silva et al. [24] presented the modeling language MAS-ML, which extends UML and was based on the conceptual framework (metamodel) called TAO [35]. The metamodel Taming Agents and Objects (TAO) provides an ontology that covers the fundamentals of Software Engineering based on agents and objects and makes possible the development of large-scale MASs [36]. This metamodel connects consolidated abstractions, such as objects and classes, and new abstractions, such as agents, roles and organizations, which are the foundations for agent-oriented software engineering. TAO presents the definition of each abstraction as a concept of its ontology and establishes the relationships between them. Fig. 3 decipts the abstractions and relationships proposed in TAO. The TAO abstractions are defined as follows:

- Object: It is a element that can be passive or reactive and has state and behavior, and can be related to other elements;
- Agent: It is an autonomous and interactive element that has a mental state. Its mental state has the following components: (i) beliefs — everything the agent knows; (ii) goals — the future states that the agent wants to achieve; (iii) plans — the sequences of actions that achieve a goal, and (iv) the actions themselves;
- Organization: It is an element that groups agents and sub-organizations, which play roles and have common goals. An organization has intra characteristics, properties and behaviours represented by the agents inside it. It may restrict the behaviour of their agents and their sub-organizations through the concept of axiom, which characterizes the global organizational constraints that agents and sub-organizations must obey;

- Object role: It is an element that guides and restricts the behaviour of an object in the organization. An object role can add information, behaviour and relationships that an object instance executes;
- Agent role: It is an element that guides and restricts the behaviour of the agent playing the role in the organization. An agent role defines (i) duties as actions that must be performed by the agent playing the role, (ii) rights as actions that can be performed by the agent playing the role, and (iii) a protocol that defines the interaction between agent roles;
- Environment: It is an element to which agents, objects and organizations belong. Environments have state and behaviour.

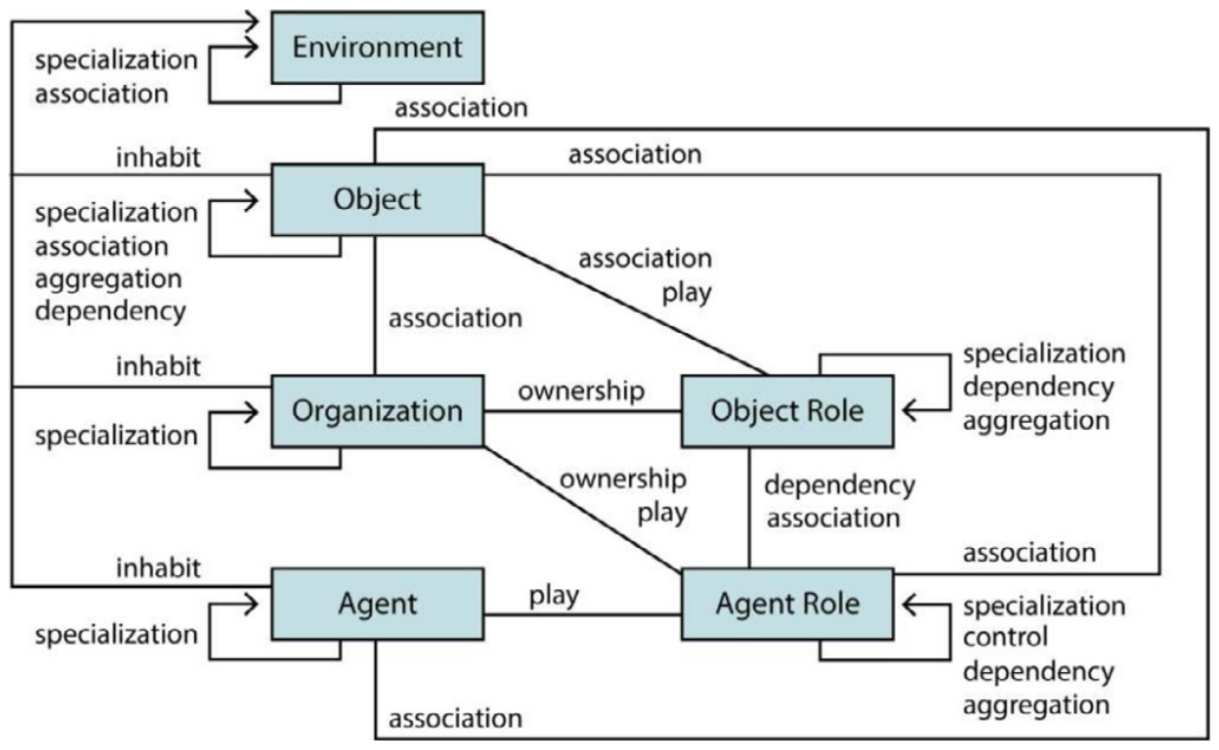

Fig. 3. Abstractions and relationships of TAO (Silva et al.,2003).

In [36], the authors define the following relationships in the TAO metamodel: Inhabit, Ownership, Play, Specialization/Inheritance, Control, Dependency, Association and Aggregation/Composition. The concepts are presented through a semi-formal approach that uses templates to formalize Objects, Agents, Organization, Object Role, Agent Role, Environment and their relationships.

MAS-ML is a modelling language that implements the TAO concepts and designed to support the modelling of agents that are goal-based in plans. MAS-ML models all structural and dynamic aspects defined in the TAO metamodel by extending the UML metamodel. The structural diagrams defined by MAS-ML are Role Diagram, Class Diagram and Organization Diagram [36]. The dynamic diagrams are Sequence Diagram and Activities Diagram [24]. Using all these diagrams, it is possible to model all structural and dynamic aspects of entities defined in TAO. Although the authors bring advances to deal with the different types of behavior found in the literature, they did not include abstractions related to norms and adaptation.There are many reasons for selecting MAS-ML for this study, namely that it enables modeling: (i) abstractions, such as agents, organization and environments; (ii) the interaction between entities; (iii) internal execution, and (iv) interaction protocols between agents.

The great difference between this approach and others is the clear definition and representation of abstractions and behaviors that make up a MAS. However, the approach in [24] is incomplete because important abstractions such as adaptation and norms are not described neither in the TAO nor in the MAS-ML modeling language.

Gonçalves and colleagues [29] proposed an extension of MAS-ML [24] called MAS-ML 2.0, which bring, as a main novelty, mechanisms that support modeling proactive agents, and also the ability to model different internal architectures of agents, such as those presented in (Russell and Norving, 2003). These different architectures were created to deal only with agent behavior based on reactive and proactive fundamentals.

The modeling language NormML [20] based on UML was developed to specify norms that restrict the behavior of entities in MASs. The NormML metamodel provides a language to model roles, permissions, actions, resources, and authorization constraints along with the relationships between permissions and roles, actions and permissions, resources and actions, and constraints and permissions. This approach allows the modeling of the static aspects of norms, but it is not possible to define dynamic aspects of a norm or to define norms in an interactive context. In addition, it is not possible to identify the rules that are active or those that have been violated; the norms described in NormML are not able to undergo changes throughout their activation time in the environment.

The work presented in [37] describes the AUML language. This modelling language aims to provide semi formal and intuitive semantics through a friendly graphical notation. AUML does not provide elements to represent the next-function, planning, formulate problem function, formulate goal function and utility function. For instance, this language does not define the environment as an abstraction, so it is not possible to model agents that are able to move from one environment to another. Further, AML [16] is a modelling language based on a metamodel that enables the modelling of organizational units, social relations, roles and role properties. AML agents are composed of attribute list, operation list, parts and behaviors, and sensors and actuators.

### C. Research Opportunities

In this section we identify some the gaps in existing multi-agent modeling approaches and compare the expressiveness of these approaches using abstract atomic elements related to agents, norms and adaptation.

The approaches described in the previous section do not support modeling concepts related to both adaptation and norms and their respective interactions [34], [38]. In addition, they have several drawbacks, which led to the choice of extending MAS-ML in order to model adaptation and norms. Nevertheless, norms and adaptation are relevant and indispensable to the internal and external design and run times of MAS modeling. Even though the literature reports several languages [14], [39], [40], [22], [15], [10] with capabilities for modeling MASs, there is still a need for a modeling language to describe concepts related to adaptation and norms as first-class abstractions through an explicit MAS metamodel; that can be used to model the structural and dynamic aspects often described in MASs for these concepts, and promote the refinement of these models from design into code.

Current organizational norms are mostly used to restrict the behavior of agents and are implemented in the following ways: (i) addressing a given role; (ii) as a set of states to be regulated by the norm which also affects the agent's role; (iii) as valid restrictions when the norm is active in the environment; (iv) in many modeling languages and methodologies using the deontic concept of obligation [17], [22], [18], and (v) through the rewards and punishments associated with a particular norm [11].

Regarding the adaptation concept, we make explicit in the metamodel atomic concepts such as: (i) monitoring sensors; (ii) effectors, which are mechanisms needed to run the agent's actions and plans; (iii) events of the specific system — for example, to show when the system objectives are violated; (iv) strategies to adapt to environmental conditions; (v) different tactics performing plans in each strategy to carry out the adaptation, and (vi) adaptation policies that should always be checked during the environmental monitoring [8], [10].

Table II presents a comparison among the expressiveness of the main multi-agent architectures and models found in the literature ( [17], [14], [20], [18], [19], [28]) based on the atomic abstractions related to the concepts of software agents, norms and adaptation. Some modeling languages and methodologies, such as AUML [37] and MAS-ML [24], Adelfe [33], FAML [14] and Gaia [27], do not support the modeling of norms. Approaches such as NormML [20] also make the modeling of several elements of a norm possible. From the set of modeling languages [40], [22], [15], and methodologies [14], [39], [10] we have reviewed, just one of them (NormML) is able to model all the properties of the elements described in Section 2.2. However, NormML does not support the representation of adaptation-related abstractions (Section 2.1) and, as a consequence, does not support modeling the interactions between norms and adaptation. In terms of abstractions for adaptation, although only Adelfe [33] is able to represent adaptive agents using adaptive workflows, it does not introduce the atomic elements an agent needs to adapt its behavior.

The presented comparison serves as a basis for the definition of a new MAS metamodeling language called ANA-ML. Note that, according to the last column in Table II, ANA-ML should be able to capture not only atomic abstractions related to agents, but also those related to adaptation and norms.

## IV. Conclusion and Prospects

In this paper we have presented a survey of existing MAS modeling approaches and explored research opportunities that arise as these systems involve abstractions such as norms and adaptation. As we have compared the expressiveness of existing MAS modeling approaches, we realized there is a gap in terms of providing a MAS modeling model and language that could support norms, adaptation and their interactions.

In the following we discus some future research prospects towards facilitating the design and development of Adaptive Normative Multi-Agent Systems (ANMASs).

### *A. Defining a Novel Adaptive Normative Agent (ANA) Metamodel*

Based on the atomic abstractions related to agents, norms and adaptation, there is a need to define an Adaptive Normative Agent (ANA) metamodel. This ANA metamodel should enable representing the structural and adaptive behavior of agent-based software to support norms. This metamodel should represent the internal and external entities and relationships of a software agent. The internal agent representation may describe the agent internal properties and their relationships. The external agent representation may define the relationships among agents and other entities in the environment.

### *B. Providing a New Adaptive Normative Agent Modeling Language (ANA-ML)*

Based on the ANA metamodel, an adaptive normative modeling language (ANA-ML) can be defined. ANA-ML should introduce adaptation and norm-related abstractions in the modeling language MAS-ML through a UML extension.

### *C. Developing Case Studies and Experimental Studies*

Case studies involving the use of ANA-ML in specific scenarios should be conducted. Potential application domains include disaster management (e.g., in case of fire or floods), social networks, security and smart cities. These cases studies can demonstrate that ANA-ML can be feasibly applied in real world applications.

Experimental studies can also be developed to assess ANA-ML. Empirical studies should evaluate ANA-ML with other modeling languages, for example, with respect to representation understanding, time and difficulty. This is relevant in the assessment of how complex it is to use ANA-ML to model specific scenarios.

## Acknowledgment

This work has been supported by the Laboratory of Software Engineering (LES) at PUC-Rio. Our thanks to CAPES, CNPq, FAPERJ, PUC-Rio for their support through scholarships and fellowships, and the Natural Sciences and Engineering Council of Canada (NSERC).

TABLE II
COMPARING THE EXPRESSIVENESS OF MULTI-AGENT MODELING APPROACHES USING ABSTRACT ATOMIC ELEMENTS RELATED TO AGENTS, NORMS AND ADAPTATION

| | | **AUML** | **AML** | **GAIA** | **ADELFE** | **FAML** | **MAS-ML** | **MAS-ML 2.0** | **NormML** | **ANA-ML** |
|---|---|---|---|---|---|---|---|---|---|---|
| Agent | Belief | | X | | X | X | X | X | | X |
| | Desire | | | | | X | | | | X |
| | Intention | | | | | X | X | X | | X |
| | Action | | X | X | X | X | X | X | | X |
| | Plan | | X | | X | X | X | X | | X |
| | Goal | | X | X | X | X | X | X | | X |
| | Role | | X | X | X | X | X | X | | X |
| Adaptive | Sensor | | | | X | | | | | X |
| | Effector | | | | X | | | | | X |
| | Event | | | | X | | | | | X |
| | Strategy | | | | X | | | | | X |
| Norm | Policies | | | | X | | | | | X |
| | Addressee | | | | | | | | X | X |
| | State | | | | | | | | X | X |
| | Condition | | | | | | | | X | X |
| | Deontic Concept | | X | X | | | | | X | X |
| | Motivation | | | | | | | | X | X |